\documentstyle[12pt]{article}
\begin{document}
 \newcommand{\be}[1]{\begin{equation}\label{#1}}
 \newcommand{\ee}{\end{equation}}
 \newcommand{\beqn}[1]{\begin{eqnarray}\label{#1}}
 \newcommand{\eeqn}{\end{eqnarray}}
\newcommand{\bd}{\begin{displaymath}}
\newcommand{\ed}{\end{displaymath}}
\newcommand{\mat}[4]{\left(\begin{array}{cc}{#1}&{#2}\\{#3}&{#4}\end{array}
\right)}
\newcommand{\matr}[9]{\left(\begin{array}{ccc}{#1}&{#2}&{#3}\\{#4}&{#5}&{#6}\\
{#7}&{#8}&{#9}\end{array}\right)}
 \newcommand{\eps}{\varepsilon}
\newcommand{\ov}{\overline}
\renewcommand{\to}{\rightarrow}
\renewcommand{\thefootnote}{\fnsymbol{footnote}}
%
%
\makeatletter
\newcounter{alphaequation}[equation]
\def\thealphaequation{\theequation\alph{alphaequation}}
%
\def\eqnsystem#1{
\def\@eqnnum{{\rm (\thealphaequation)}}
\def\@@eqncr{\let\@tempa\relax
\ifcase\@eqcnt \def\@tempa{& & &}
\or \def\@tempa{& &}\or \def\@tempa{&}\fi\@tempa
\if@eqnsw\@eqnnum\refstepcounter{alphaequation}\fi
\global\@eqnswtrue\global\@eqcnt=0\cr}
\refstepcounter{equation}
\let\@currentlabel\theequation
\def\@tempb{#1}
\ifx\@tempb\empty\else\label{#1}\fi
\refstepcounter{alphaequation}
\let\@currentlabel\thealphaequation
\global\@eqnswtrue\global\@eqcnt=0
\tabskip\@centering\let\\=\@eqncr
$$\halign to \displaywidth\bgroup
  \@eqnsel\hskip\@centering
  $\displaystyle\tabskip\z@{##}$&\global\@eqcnt\@ne
  \hskip2\arraycolsep\hfil${##}$\hfil&
  \global\@eqcnt\tw@\hskip2\arraycolsep
  $\displaystyle\tabskip\z@{##}$\hfil
  \tabskip\@centering&\llap{##}\tabskip\z@\cr}

\def\endeqnsystem{\@@eqncr\egroup$$\global\@ignoretrue}
\makeatother

\begin{titlepage}
\thispagestyle{empty}

\begin{flushright}
INFN-FE 02-95 \\
hep-ph/9503366 \\
February 1995
\end{flushright}
\vspace{15mm}

\begin{center}
{\Large \bf SUSY SU(6) GIFT for \\
Doublet-Triplet Splitting and Fermion Masses } \\
\vspace{1.0cm}
{\large Zurab Berezhiani} \\
\vspace{5mm}
{\it INFN Sezione di Ferrara, I- 44100 Ferrara, Italy } \\
{\it Institute of Physics, Georgian Academy of Sciences,
380077 Tbilisi, Georgia  }
\end{center}

\vspace{1.0cm}
\begin{abstract}
The supersymmetric $SU(6)$ model equipped by the
flavour-blind discrete gauge symmetry $Z_3$ is considered.
It provides simultaneous solution to the
doublet-triplet splitting problem, $\mu$-problem and leads
to natural understanding of fermion flavour.
The Higgs doublets arise as Goldstone modes
of the spontaneously broken
{\em accidental} global $SU(6)\times U(6)$ symmetry of the
Higgs superpotential. Their couplings to fermions have
peculiarities leading to the consistent picture of the quark and
lepton masses and mixing, without invoking the
horizontal symmetry or zero texture concepts. In particular,
the only particle that has direct $O(1)$ Yukawa coupling with the
Higgs doublet is top quark. Other fermion masses arise
from the higher order operators, with natural
mass hierarchy described in terms of small ratios $\eps_\Sigma=
V_\Sigma/V_H$ and $\eps_H=V_H/M$, where $V_H$ and $V_\Sigma$
correspondingly are the $SU(6)$ and $ SU(5)$ symmetry breaking
scales, and $M$ is a large (Planck or string) scale.
 The model automatically implies almost precise
$b-\tau$ Yukawa unification. Specific mass formulas are also obtained,
relating the down quark and charged lepton masses.
Neutrinos get small ($\sim 10^{-5}\,$eV) masses
which can be relevant for solving the solar
neutrino problem via long wavelength vacuum oscillation.
\end{abstract}

\vfil
\end{titlepage}

\rm\baselineskip=14pt






\renewcommand{\thefootnote}{\arabic{footnote})}
\setcounter{footnote}{0}

\newpage

{\large \bf 1. Introduction }
\vspace{0.5cm}

The evidence of the gauge coupling unification \cite{crossing} in the
minimal supersymmetric standard model (MSSM) suggests the following
paradigm: at the Planck or string scale $M\sim 10^{18-19}\,$GeV the
ultimate ``Theory of Everything" reduces to a field theory given by the
SUSY GUT, which then is broken at the scale $M_X\simeq 10^{16}\,$GeV
down to the $SU(3)\times SU(2)\times U(1)$ MSSM, with minimal content
of chiral superfields including the standard fermion families and
the Higgs doublets $h_{1,2}$.

The central question dubbed a gauge hierarchy problem concerns the
origin of scales: why the electroweak scale $M_W$ is so small as compared
to the GUT scale $M_X$, which in itself is not far from the Planck scale?
It is well known \cite{Maiani} that supersymmetry can stabilize the
Higgs mass ($\sim M_W$) against radiative corrections, provided that
the soft SUSY breaking scale $m$ (typically given by the gaugino and
sfermion masses) does not exceed few TeV.  Most likely, the electroweak
scale $M_W$ emerges from the SUSY scale $m$ itself.  In particular,
it is suggestive to think that the MSSM Higgs doublets $h_{1,2}$
would stay massless in the exact SUSY limit, and the {\em only} source
of their non-zero masses is related to the soft SUSY breaking terms.
However, in the context of grand unification the gauge hierarchy problem
has the following puzzling aspects:

$A$. The problem of the doublet-triplet (DT) splitting \cite{Maiani}:
the Higgs doublets should stay light, while their colour triplet partners
in GUT supermultiplet should have $O(M_X)$ mass. Otherwise the latter would
cause unacceptably fast proton decay (mainly via the Higgsino mediated
$d=5$ operators \cite{dim5}), and also
spoil the gauge coupling unification.

$B$. The $\mu$-problem \cite{mu}: the resulting low energy MSSM should
contain the supersymmetric $\mu h_1 h_2$ term defining the higgsino
masses, with $\mu \sim M_W$. It is questionable why the supersymmetric
mass $\mu$ should be of the order of soft SUSY breaking mass $m$.

Another theoretical weakness of SUSY GUTs is a lack in the
understanding of flavour. Although GUTs can potentially unify
the Yukawa couplings within each fermion family, the origin of
inter-family hierarchy and weak mixing pattern remains open.
Moreover, for the light families the Yukawa unification simply
contradicts to the observed mass pattern, though the $b-\tau$ Yukawa
unification may constitute a case of partial but significant success.
In order to deal with the flavour problem in GUT frameworks,
some additional ideas (horizontal symmetry, specific textures)
have to be invoked \cite{Fritzsch,DHR}.

An attractive possibility towards the solution of these problems is
suggested by the GIFT ({\em Goldstones Instead of Fine Tuning})
mechanism in SUSY $SU(6)$ model \cite{BD,BDM,BDSBH},
which is a minimal extension of $SU(5)$:\footnote{The
Goldstone boson mechanism for the DT splitting was first suggested
in the context of SUSY $SU(5)$ \cite{Inoue,Anselm} (in \cite{Anselm}
it was elegantly named as GIFT), by assuming an {\em ad hoc} $SU(6)$
global symmetry of the Higgs superpotential. Our results, however,
are specific of the gauged $SU(6)$ theory. }
the Higgs sector contains supermultiplets $\Sigma$ and
$H+\bar{H}$ respectively in adjoint 35 and fundamental $6+\bar{6}$
representations, in analogy to 24 and $5+\bar{5}$ of $SU(5)$.
However, this model drastically differs from the other GUT approaches.
Usually in GUTs the Higgs sector consists of two different sets:
one is for the GUT symmetry breaking (e.g. 24-plet in $SU(5)$),
while another containing the Higgs doublets (like $5+\bar{5}$ in $SU(5)$)
is just for the electroweak symmetry breaking and fermion mass generation.
In contrast, the $SU(6)$ theory has no special superfields for the second
function: 35 and $6+\bar 6$ constitute a minimal Higgs content
needed for the local $SU(6)$ symmetry breaking down to MSSM.\footnote{
In order to maintain the gauge coupling unification, $SU(6)$ must be first
broken to $SU(5)$ by $H,\bar H$ at some scale $V_H$.
At this stage the fermion sector is also reduced \cite{BD,BDSBH}
to the minimal $SU(5)$ content.
Then at the scale $V_\Sigma\simeq 10^{16}\,$GeV, $\Sigma$ breaks the
intermediate $SU(5)$ down to $SU(3)\times SU(2)\times U(1)$. }
As for the light Higgs doublets $h_{1,2}$, they arise from the doublet
fragments in $\Sigma$ and $H,\bar{H}$, as Goldstone modes of the
accidental global symmetry $SU(6)_\Sigma \times U(6)_H$.
This global symmetry arises \cite{BD} if mixing terms of
the form $\bar{H}\Sigma H$ are suppressed in the Higgs superpotential.
Thus $h_{1,2}$ being strictly massless in the exact SUSY limit, acquire
non-zero mass terms (including the $\mu$-term) {\em only} due to
the spontaneous SUSY breaking and subsequent radiative corrections.

On the other hand, in the GIFT picture the Yukawa couplings have
peculiarities leading to new possibilities towards the understanding of
flavour. Indeed, if the Yukawa terms also respect the
$SU(6)_\Sigma \times U(6)_H$ global symmetry, then $h_1$ and $h_2$ being
the Goldstone modes should have {\em vanishing} Yukawa couplings with
the fermions that remain massless after the $SU(6)$ symmetry breaking
down to MSSM, that are ordinary quarks and leptons.
Thus, the couplings relevant for fermion masses should {\em explicitly}
violate $SU(6)_\Sigma \times U(6)_H$. This constraint leads to striking
predictions for the fermion mass and mixing pattern even in completely
`democratic' approach, without invoking the horizontal symmetry arguments.
In particular, it was shown in \cite{BDSBH} that {\em only} the top quark
can get $\sim 100\,$GeV mass through renormalizable $SU(6)$ invariant
Yukawa coupling. For the other fermion masses one has to appeal to the
higher order operators, scaled by inverse powers of the Planck scale.
In order to achieve a proper operator structure, additional discrete
symmetry was invoked. The model suggested in \cite{BDSBH} succeeded in
appealing description of the third and second fermion families,
but the first family was rendered massless.

In order to built a consistent GIFT model, one has to find some valid
symmetry reasons to forbid the mixing terms like $\bar{H}\Sigma H$:
otherwise the theory has no accidental global symmetry.
It is natural to use for this purpose the discrete gauge symmetries,
which can naturally emerge in the string theory context.
In the present paper we suggest a consistent SUSY $SU(6)$ model
equipped with the flavour-blind discrete $Z_3$ symmetry.
The role of the latter is important:
it forbids the mixing terms in the Higgs superpotential thus
ensuring the accidental $SU(6)_\Sigma \times U(6)_H$ symmetry,
and provides the proper higher order operators for
generating a realistic mass and mixing pattern of {\em all} fermions.

\vspace{0.9cm}
{\large \bf 2. $SU(6)\times Z_3$ model }
\vspace{0.5cm}

Let us assume that below the Planck or string scale $M$ the theory
is given by SUSY GUT with the $SU(6)$ gauge symmetry, containing the
following chiral superfields --
`Higgs' sector: vectorlike supermultiplets $\Sigma_1(35)$,
$\Sigma_2(35)$, $H(6)$, $\bar{H}(\bar{6})$ and an auxiliary
singlet $Y$;
`fermion' sector: chiral, anomaly free supermultiplets
$(\bar{6} + \bar{6}^\prime)_i$, $15_i$ ($i=1,2,3$ is a family index)
and 20;  and
some heavy vector-like matter multiplets like $15_F+\ov{15}_F$,
etc., which we recall later on as $F$-fermions.
According to survival hypothesis \cite{surv}, these should have
$SU(6)$ invariant large ($\sim M$) mass terms and thus decouple from
the lighter sector.\footnote{
The survival hypothesis does not apply to 20, since it is a pseudo-real
representation and the mass term $M\,20\,20$ is vanishing (the singlet
is contained only in antisymmetric tensor product $20\times 20$).
More generally, if in the original theory $20$-plets present in odd
number then {\em one} of them inevitably `survives' to be massless. }
However, they can play a crucial role in the light fermion mass
generation \cite{Frogatt}. In Sect. 4 we use the
$F$-fermion exchanges for inducing the masses of all light fermions,
except the top which gets mass from the direct Yukawa coupling.

We introduce also two flavour-blind discrete symmetries.
One is usual matter parity $Z_2$, under which
the fermion superfields change the sign while the Higgs ones stay
invariant. Such a matter parity, equivalent to R parity,
ensures the proton stability.
Another discrete symmetry is $Z_3$ acting in the following way
$(\omega=\mbox{e}^{{\rm i}\frac{2\pi}{3}})$:
\be{Z3}
  20 \to \omega\, 20, ~~~ 15_i \to \bar{\omega}\, 15_i, ~~~
\bar{6}_i,\bar{6}'_i \to \omega\, \bar{6}_i,\bar{6}'_i, ~~~~
\Sigma_1 \to \omega\, \Sigma_1, ~~~
\Sigma_2 \to \bar{\omega}\,\Sigma_2,
\ee
while $H,\bar{H}$ and $Y$ are invariant. One can easily check that
this $Z_3$ symmetry satisfies the anomaly cancellation constraints
\cite{Ibanez} so that it can be regarded as the gauge discrete symmetry.
The matter parity $Z_2$ is also known to be free of discrete anomalies
\cite{Ibanez}.

Let us consider first the Higgs sector.
The most general renormalizable superpotential compatible with the
$SU(6)\times Z_3$ symmetry is\footnote{ we assume
that all coupling constants
are of the order of 1, say within factor of $3-4$.
For comparison, we remind that the gauge coupling constant
at the GUT scale is $g_X\simeq 0.7$ }
\be{superpot}
W = M_\Sigma\Sigma_1 \Sigma_2 + \lambda_1 \Sigma_1^3 + \lambda_2 \Sigma_2^3
+ \lambda S \Sigma_1 \Sigma_2
+ M_H \bar{H}H + \rho Y (\bar{H}H - \Lambda^2) +
M_Y Y^2 + \xi Y^3
\ee
This superpotential automatically has the global symmetry
$SU(6)_{\Sigma}\times U(6)_H$, related to independent transformations
of $\Sigma$ and $H$.\footnote{ In fact,
$SU(6)_\Sigma\times U(6)_H$ is not a global symmetry of a whole
Lagrangian, but only of the Higgs superpotential. In particular, the
Yukawa as well as the gauge couplings ($D$-terms) do not respect it.
However, in the exact supersymmetry limit (i) it is
effective for the field configurations on the vacuum valley,
where $D=0$, (ii) owing to non-renormalization theorem, it cannot be
spoiled by the radiative corrections from the Yukawa interactions. }
In the exact SUSY limit
the condition of vanishing $F$ and $D$ terms allows, among the other
degenerated vacua, VEVs\footnote{Discrete degeneration
of the $\langle \Sigma \rangle$ is not essential and will be
immediately removed for the proper range of the soft SUSY breaking
parameters $A,B$ (see below, eq. (\ref{SB_terms})). However, for
$\langle H\rangle,\langle\bar{H} \rangle$ fixed as in eq. (\ref{VEVs})
there is also continuous degeneration related to independent rotation
of $\langle \Sigma \rangle$: any configuration obtained by the
unitary transformation $U^\dagger \langle \Sigma_{1,2} \rangle U$
is a vacuum state as well. Actually this flat direction gives rise
to Goldstone mode which can be identified to the Higgs doublets
provided that true vacuum is given by $U=1$, i.e. the relative
orientation of the VEVs is fixed as in eq. (\ref{VEVs}).
For a proper parameter range, this configuration can indeed appear
as a true vacuum state after lifting the vacuum degeneracy by the effects
of SUSY breaking and subsequent  radiative corrections \cite{BDM}. }
\be{VEVs}
\langle \Sigma_{1,2} \rangle = V_{1,2} \left( \begin{array}{cccccc}
1 & & & & & \\& 1 & & & & \\ & & 1 & & & \\ & & & 1 & & \\ & & & & -2 & \\
& & & & & -2 \end{array} \right), ~~~~
\langle H \rangle =\langle \bar{H} \rangle = V_H \left( \begin{array}{c}
1\\0\\0\\0\\0\\0 \end{array} \right),~~~
\langle Y \rangle= V_Y
\ee
where, provided that $\Lambda\gg V_\Sigma=(V_1^2+V_2^2)^{\frac{1}{2}}$,
we have:
\be{VEV}
V_Y = \frac{M_H}{\rho},~~~~
V_{1,2}=\frac{ M_\Sigma + \lambda V_Y }
{ (\lambda_1\lambda_2)^{\frac{1}{3}}\lambda_{1,2}^{\frac{1}{3}}} \,,~~~~
V_H=\Lambda + O\left(\frac{V_\Sigma^2}{\Lambda}\right)
\ee
These VEVs lead to needed pattern of the gauge symmetry breaking:
$H,\bar{H}$ break $SU(6)$ down to $SU(5)$, while $\Sigma_{1,2}$ break
$SU(6)$ down to $SU(4)\times SU(2)\times U(1)$.
Both channels together break the local symmetry down to
$SU(3)\times SU(2)\times U(1)$.
At the same time, the global symmetry $SU(6)_{\Sigma}\times U(6)_H$
is broken down to $[SU(4)\times SU(2)\times U(1)]_\Sigma \times U(5)_H$.
The Goldstone degrees which survive from being eaten by the $SU(6)$
gauge superfields via the Higgs mechanism, constitute a couple
of the MSSM Higgs doublets $h_1+h_2$ which in terms of the doublet
(anti-doublet) fragments in $\Sigma_{1,2}$ and $H,\bar{H}$ are given as
\be{Higgs}
h_2= c_\eta(c_\sigma h_{\Sigma_1} + s_\sigma h_{\Sigma_2}) -
s_\eta h_H\,,~~~~~
h_1= c_\eta(c_\sigma \bar{h}_{\Sigma_1} +
s_\sigma\bar{h}_{\Sigma_2}) - s_\eta \bar{h}_{\bar{H}}
\ee
(here and in the following we use notations
$c_\sigma=\cos\sigma$, $s_\sigma=\sin\sigma$, etc.),
where $\tan\eta=3V_\Sigma/V_H$ and $\tan\sigma=V_2/V_1=
(\lambda_1/\lambda_2)^{\frac{1}{3}}$.
In the natural range of constants $\lambda_{1,2}$ allowed to deviate
from 1 no more than a factor of $4$, $\tan\sigma \simeq 1$ within
a factor of 2.

After the SUSY breaking enters the game (presumably through the
hidden supergravity sector), the Higgs potential, in addition
to the (supersymmetric) squared $F$ and $D$ terms, includes also
the soft SUSY breaking terms \cite{BFS}.
These are
\be{SB_terms}
V_{SB}=AmW_3 + BmW_2 + m^2\sum_k|\phi_k|^2,
\ee
where $\phi_k$ imply all scalar fields involved, $W_{3,2}$
are terms in superpotential respectively trilinear and bilinear
in $\phi_k$, and $A,B,m$ are soft breaking parameters.
Due to these terms the VEVs $V_{1,2}$ are shifted by an amount of
$\sim m$ as compared to the ones in eq. (\ref{VEV}) being calculated
in the exact SUSY limit. Via the $\Sigma^3$ terms in the superpotential,
this shift gives rise to term $\mu h_1 h_2$ contributing
the higgsino masses. Thus, the GIFT scenario automatically solves the
$\mu$-problem: the (supersymmetric) $\mu$-term for the resulting MSSM
in fact arises in consequence of SUSY breaking, with $\mu\sim m$.

The scalar components of $h_{1,2}$ acquire the soft SUSY breaking mass
terms, but not all of them immediately. Clearly, $V_{SB}$ also
respects the larger global symmetry $SU(6)_\Sigma\times U(6)_H$,
so that only the combination $h=h_1-h_2^\ast$ of scalars gets a
$\sim m$ mass, while the orthogonal state $\tilde{h}=h_1+h_2^\ast$
remains massless as a truly Goldstone boson. Taking into the account
also the structure of $D$-term, we see that there is a vacuum valley
with $v_2/v_1=1$, where $v_{1,2}$ are the VEVs of $h_{1,2}$
while the value of the $v_1=v_2$ remains arbitrary.

However, SUSY breaking relaxes radiative corrections
(mainly due to the large top Yukawa coupling)
which lift the vacuum degeneracy and provide non-zero mass to
$\tilde{h}$, fixing thereby the VEVs $v_1$ and $v_2$.
It is natural to expect that renormalization effects will not deviate
these VEVs very strongly from the valley given by $v_1=v_2$, so that
the magnitude of $\tan\beta=v_2/v_1$ will be very moderate.
The effects of radiative corrections leading to the
electroweak symmetry breaking were studied in ref. \cite{BDM}.
It was shown that in spite of earlier claims \cite{Anselm,Goto}
the GIFT scenario does not imply any upper bound
on the top mass,  and it can go up to its infrared
fixed limit $M_t=(190-210)\sin\beta\,$GeV \cite{IRfixed}.

Thus, our model naturally solves both the DT splitting and the
$\mu$ problems. The Higgs doublets $h_{1,2}$ remain light,
while their triplet partners are superheavy.
Indeed, the triplet fragments from $\Sigma_{1,2}$ have masses
$\sim V_\Sigma$,  and the triplets from $H, \bar H$ are the Goldstone
modes eaten up by the $SU(6)$ gauge superfields.
%
In the following  we assume that $V_H\gg V_\Sigma$, as it is suggested
by the gauge coupling unification, and show how the observed hierarchy of
fermion masses can be naturally explained in terms of small ratios
$\eps_\Sigma=V_\Sigma/V_H$ and $\eps_H=V_H/M$.
In this case the Higgs doublets dominantly come from $\Sigma_{1,2}$
while in $H,\bar{H}$ they are contained with small
weight $\sim 3\eps_H$.

\vspace{0.9cm}
\newpage
{\large \bf 3. Fermion masses: general operator analysis }
\vspace{0.5cm}

The most general Yukawa superpotential allowed by the
$SU(6)\times Z_3$ symmetry is
\be{Yukawa}
W_{Yuk} = G\,20 \Sigma_1 20 \,+ \,\Gamma\,20 H 15_3\,  + \,
\Gamma_{ij} 15_i \bar{H} \bar{6}^\prime_j\,, ~~~~~~~~i,j=1,2,3
\ee
where all Yukawa coupling constants are assumed to be $O(1)$.
Without loss of generality, one can always redefine the basis of 15-plets
so that only the $15_3$ state couples 20-plet in (\ref{Yukawa}).
Also, among six $\bar{6}$-plets one can always choose three of them
(denoted in eq. (\ref{Yukawa}) as $\bar{6}^\prime_{1,2,3}$) which couple
$15_{1,2,3}$, while the other three states $\bar{6}_{1,2,3}$ have
no Yukawa couplings.

Already at the scale $V_H$ of the gauge symmetry breaking
$SU(6) \to SU(5)$
the fermion content of our theory reduces to the one of minimal $SU(5)$.
Indeed, the $SU(5)\supset SU(3)\times SU(2)\times U(1)$ decomposition of
the fermion multiplets under consideration reads
\beqn{fragments}
& & 20=10 + \ov{10} = (q+u^c+e^c)_{10}+(Q^c+U+E)_{\ov{10}} \nonumber \\
& & 15_i=(10+5)_i = (q_i+u^c_i+e^c_i)_{10} +
(D_i + L^c_i)_5  \nonumber \\
& & \bar{6}_i=(\bar{5}+1)_i = (d^c_i + l_i)_{\bar{5}} + n_i \nonumber \\
& & \bar{6}_i^\prime=(\bar{5}+1)_i^\prime =
(D^c_i +L_i)_{\bar{5}^\prime} + n_i^\prime\,,~~~~~~~~~i=1,2,3
\eeqn
According to eq. (\ref{Yukawa}), the extra fermion pieces with
non-standard $SU(5)$ content, namely $\ov{10}$ and $5_{1,2,3}$,
form massive particles being coupled to $10_3$ and $\ov{5}'_{1,2,3}$:
\be{heavymass}
\Gamma\,V_H\,\ov{10}\,10_3\, +
\,\Gamma_{ij}V_H\,5_i\,\bar{5}_j^\prime\, +
\,G\, V_1 \,(U\,u^c - 2 E\,e^c) \,,
\ee
and thereby decouple from the light states which remain as
$\bar{5}_{1,2,3}$, $10_{1,2}$ and 10 (we neglect the small
($\sim \eps_\Sigma$) mixing between the $u^c - u^c_3$ and $e^c - e^c_3$
states) and singlets $n_i,n'_i$.

The couplings of 20-plet in (\ref{Yukawa}) explicitly
violate the global $SU(6)_\Sigma\times U(6)_H$ symmetry. Hence,
the up-type quark from 20 (to be identified as top)
has {\em non-vanishing} coupling with the Higgs doublet $h_2$.
As far as $V_H\gg V_\Sigma$, it essentially emerges
from $G\,20\Sigma_1 20 \to G\, q u^c \,h_2.\,$
Thus, in our scheme {\em only} the top quark can have $\sim 100\,$GeV
mass due to the large Yukawa constant $\lambda_t=G \sim 1$.
Other fermions would stay massless unless we invoke the higher
order operators scaled by inverse powers of the large mass $M$.
Such operators could appear due to quantum gravity effects,
with $M\sim M_{Pl}$. Alternatively, they can arise by integrating
out heavy fermions with masses $M\gg V_H$ (see Sect. 4).

Nevertheless, before addressing the concrete scheme with heavy
fermion superfields, let us start with the general operator analysis.
Obviously, $Z_3$ symmetry forbids the $d=5$ `Yukawa' terms in the
superpotential.\footnote{ Operators involving an odd number of
fermion superfields are forbidden by $Z_2$ matter parity.}
However, the $d=6$ operators are allowed and they
are the following:\footnote{The way of the $SU(6)$ indices
convolution in these operators is indicated by the parentheses
so that the combinations inside  transform as effective $\bar 6$ or $6$.
We remind that operators which are relevant for the light fermion masses
should {\em explicitly} violate the global $SU(6)_\Sigma\times U(6)_H$
symmetry. The possible terms $15 \bar H (\Sigma_1\Sigma_2 \bar 6)$ and
$15 \bar H \bar 6\cdot {\rm Tr}(\Sigma_1\Sigma_2)$ actually do
not violate it and therefore are irrelevant. We also omit the operators
obtained by trivial replacing $\Sigma_1\to \Sigma_2$ in ${\cal S}$. }
\be{dim4_btau}
{\cal B} = \frac{B}{M^2}\, 20 \bar{H} (\Sigma_1 \bar{H}) \bar{6}_3\,,~~~~~
{\cal C} = \frac{ C_{ij} }{M^2}\, 15_i H (\Sigma_2 H) 15_j
\ee
\be{dim4_smu}
{\cal S} = \frac{S^{(1)}_{ik} }{M^2}\, 15_i(\Sigma_1\Sigma_2\bar{H})
\bar{6}_k +
\frac{S^{(2)}_{ik} }{M^2}\,15_i(\Sigma_1 \bar{H})(\Sigma_2 \bar{6}_k)
\ee
\be{dim4_nu}
{\cal N} = \frac{N_{kl} }{M^2}\, \bar{6}_k H (\Sigma_1 H) \bar{6}_l
\ee
(clearly, matrices $C_{ij}$ and $N_{kl}$ are symmetric) where
$B,\dots N_{kl}~$ are the $O(1)$ constants.

First we focus on the operators ${\cal B}$, ${\cal C}$ and ${\cal S}$
generating the charged fermion masses. (${\cal N}$ is relevant only
for the neutrino masses, and we consider it later in this section).
Similar operators involving heavy $\bar{6}^\prime_i$ states are
irrelevant, since the charged fragments of the latter are already massive.
According to eq. (\ref{heavymass}), the state
$10_3\subset 15_3$ is also heavy and it is decoupled from the light
particle spectrum. Therefore, these operators are relevant
only for $10\subset 20$,
$10_i\subset 15_i$ ($i=1,2$) and $\bar{5}_k\subset \bar{6}_k$
($k=1,2,3$) states. Without loss of generality, we redefine the basis
of $\bar{6}$-plets so that only the $\bar{6}_3$ state couples
20 in eq. (\ref{dim4_btau}).

Obviously, the operator ${\cal B}$ is responsible for the $b$ quark
and $\tau$ lepton masses, and at the MSSM level it reduces to the Yukawa
couplings $\eps_H^2 c_\sigma B\,(q d^c_3  + e^c l_3)\,h_1$.
Hence, though $b$ and $\tau$ belong to the same
family as $t$ (namely, to 20-plet), their Yukawa constants are
substantially (by factor $\sim \eps_H^2$) smaller than $\lambda_t$.
Moreover, we automatically have almost precise $b-\tau$ Yukawa
unification at the GUT scale:
\be{Yuk_btau}
\lambda_b=\eps_H^2 c_\sigma B\,,~~~~~~
\lambda_\tau= \eps_H^2 c_\sigma B\,[1-\eps_\Sigma^2(c_\sigma G/\Gamma)^2]
\cong \lambda_b
\ee
where the $\sim \eps_\Sigma^2$ correction is due to the mixing of
$e^c$ and $e^c_3$ states in eq. (\ref{heavymass}).

As far as the third family fermions are already defined as the states
belonging to $20$ and $\bar{6}_3$, operators ${\cal C}$ and ${\cal S}$
induce mass terms for the fermions of the first two families, which
in general would appear unsplit. Indeed, for the
Yukawa matrices of the corresponding upper and down quarks
and charged leptons we obtain:
\be{Yuk_mat}
\lambda_{ij}^{\rm up}=\eps_H^2 s_\sigma C_{ij}\,,~~~~
\lambda_{ik}^{\rm down}=\eps_\Sigma\eps_H^2 s_\sigma c_\sigma
(S^{(1)}_{ik} - S^{(2)}_{ik})\,,~~~~
\lambda_{ik}^{\rm lept}=\eps_\Sigma\eps_H^2 s_\sigma c_\sigma
(S^{(1)}_{ik} + 2S^{(2)}_{ik})
\ee
Thus, for $\eps_H,\eps_\Sigma\sim 0.1$ a feasible description
of the third and second family masses can be achieved:
we naturally
(without appealing to any flavour symmetry) obtain
$\lambda_t \gg \lambda_{\tau(b)},\lambda_c \gg \lambda_{s,\mu}$.
The charm quark Yukawa constant $\lambda_c\sim \eps_H^2$,
as well as the bottom-tau constant (\ref{Yuk_btau}), whereas the
$\lambda_{s,\mu}$ are smaller by factor of $\sim\eps_\Sigma$.\footnote{
As we have commented earlier, the natural value of $\tan\sigma$ is
of about 1.
The fact that the physical masses of $b$, $\tau$ and $c$ are all
in the GeV range hints that $\tan\beta$ should be close to 1,
in agreement with our earlier remark that
the natural value of $\tan\beta$ in the GIFT scenario should be very
moderate. }
In addition, the Yukawa couplings $\lambda_s$ and $\lambda_\mu$
are split due to different contribution of the second term in
(\ref{dim4_smu}).
Finally, the operator ${\cal S}$ involving the $\bar{6}_3$ state
gives rise to the $O(\lambda_s/\lambda_b)$ CKM mixing angle between the
second and third families.

However, a completely general operator analysis implies
that $\lambda_u\sim\lambda_c$ and $\lambda_{d,e}\sim \lambda_{s,\mu}$.
In order to explain the observed mass hierarchy between the first and
the second families,
some additional ideas are needed.
For example, one can assume that the `Yukawa' matrices
$C_{ij}$ and $S^{(1,2)}_{ik}$ are {\em rank-1} matrices and
in addition $S^{(1,2)}_{ik}$ are alligned,  so that these operators
provide only {\em one} non-zero mass eigenvalue for each type of
charged fermions.
Then, without loss of generality, we can redefine the basis of
$15_{1,2}$ and $\bar{6}_{1,2}$ states so that
\be{product}
C_{ij}=\mat{0}{0}{0}{C}\,, ~~~~~
S^{(1,2)}_{ik} \propto
\left(\begin{array}{ccc}{0}&{s_\theta S_2 }&{s_\theta S_3 }\\
{0}&{c_\theta S_2 }&{c_\theta S_3 }\end{array}\right)
\ee
Hence, in this basis only $C_{22}=C$ component of the matrix $C_{ij}$
is nonzero, and $c$ quark should be identified as an up-quark state from
$q_2,u^c_2 \subset 15_2$. Then $s$ and $\mu$ are the down quark and
charged lepton states contained in $q'_2\subset 15'_2$ and
$d^c_2 \subset \bar{6}_2$, where
$15'_2= s_\theta 15_1 + c_\theta 15_2$ is an effective
combination which couples $\bar{6}_2$ and
$\bar{6}_3$ states (it is not difficult to recognize
that in fact $\theta$ is the Cabibbo angle).
In this way operators ${\cal C}$ and ${\cal S}$
provide masses of $c$, $s$ and $\mu$, rendering the $u$, $d$ and $e$
states massless.
Then for the latter one can appeal to the $d=7$ operators
($15'_1$ is defined as a state orthogonal to $15'_2$):
\be{dim5_de}
{\cal D}=\frac{D^{(1)}_{ik}}{M^3} 15'_i (\Sigma_{1}^3 \bar{H})\bar{6}_k
+ \frac{D^{(2)}_{ik}}{M^3} 15'_i (\Sigma_{1}^2 \bar{H})
(\Sigma_1 \bar{6}_k) + \frac{D^{(3)}_{ik}}{M^3} 15'_i (\Sigma_{1} \bar{H})
(\Sigma_1^2 \bar{6}_k) +
\frac{D^{(4)}_{ik}}{M^3} 15'_i (\Sigma_{1} \bar{H}) \bar{6}_k Tr(\Sigma_1^2)
\ee
\be{dim5_u}
{\cal U} = \frac{U^{(1)}_{ij}}{M^3} 15_i H (\Sigma_1^2 H) 15_j +
\frac{U^{(2)}_{ij}}{M^3} 15_i H (\Sigma_1 H)(\Sigma_1 15_j)
\ee
Operator ${\cal D}$ induces the following Yukawa couplings at the
GUT scale:
\beqn{Yuk_de}
&&\tilde{\lambda}^{\rm down}_{ik} = \eps_\Sigma^2\eps_H^3 c_\sigma^3 \,
(3D_{ik}^{(1)} - D_{ik}^{(2)} + D_{ik}^{(3)} + 12D_{ik}^{(4)}) \nonumber \\
&&\tilde{\lambda}^{\rm lept}_{ik} = \eps_\Sigma^2\eps_H^3 c_\sigma^3 \,
(3D_{ik}^{(1)} + 2D_{ik}^{(2)} + 4D_{ik}^{(3)} + 12D_{ik}^{(4)})
\eeqn
which provide $\lambda_{d,e}$ in the proper range
when $\eps_\Sigma,\eps_H \sim 0.1$.
As for the operator ${\cal U}$, for $U_{11}\sim 1$ it would lead to
$\lambda_u \sim \eps_\Sigma\eps_H^3 c_\sigma^2$, which is
parametrically one order of magnitude larger then $\lambda_d$.
It is more suggestive to assume that the matrices $U_{ij}^{(1,2)}$ have a
Fritzsch-like structure \cite{Fritzsch},
with $U^{(1,2)}_{11}=0$. Then the above estimate holds rather for
$(\lambda_u\lambda_c)^{1/2}$, and we obtain the appealing estimate
$\lambda_u \sim \eps_\Sigma^2\eps_H^4 c_\sigma^4 s_\sigma^{-1}$.
As we show in sect. 4, this pattern of the Yukawa couplings can
be indeed obtained in the heavy fermion exchange scheme.

Let us conclude this section by considering the neutrino mass pattern.
After the GUT symmetry breaking the operator (\ref{dim4_nu}) reduces
to the following terms:
\be{nu}
\frac{N_{kl}}{M^2}\,c_\sigma [V_H^2 V_\Sigma n_k n_l +
V_H^2 (l_k n_l + l_l n_k) h_2 - 3V_\Sigma l_k l_l h_2^2 ]
\ee
It is not difficult to recognize in this pattern the well-known
`seesaw' picture for the neutrino mass generation.
Indeed, the `right-handed' neutrinos $n_k$ acquire large
($\sim \eps_H^2 V_\Sigma$) Majorana masses, while the second term in
eq. (\ref{nu}) is nothing but Dirac mass terms $\sim \eps_H^2 v_2$
obtained after substituting the VEV $\langle h_2 \rangle$.
As a result of the seesaw mixing, small Majorana masses are induced
for the ordinary neutrino states $\nu_k\subset l_k$:
\be{nu_mass}
m^\nu_{kl}=\frac{\eps_H}{\eps_\Sigma}\,
\frac{N_{kl}}{M}\,c_\sigma v_2^2
\ee
Thus, for $\eps_\Sigma,\eps_H\sim 0.1$ the neutrino
masses are in the range
$m_\nu \sim M_W^2/M \sim 10^{-5}\,$eV. (Notice, that the same estimate
follows in Standard Model or $SU(5)$ with possible gravity induced
non-renormalizable operators $\frac{1}{M} l l hh$ \cite{BEG}.)
It is well-known that this mass range together with large neutrino
mixing angles, also naturally implied in our `democratic' approach
with $N_{kl}\sim 1$,
can provide a feasible solution to the solar neutrino problem
through the long wavelength ``just-so" neutrino oscillations
(for recent discussions of the experimental status
of this scenario see \cite{Petcov}).

\vspace{0.9cm}
\newpage
{\large \bf  4. Yukawa couplings from heavy particle exchanges}
\vspace{5mm}

{}From the previous section, we are left with the problem how to
split the masses of the first two families (eq. (\ref{product})
for the coupling constants in ${\cal C}$ and
${\cal S}$ was assumed {\em ad hoc}).
Now we show that this problem can be solved, still without
appealing to any flavour symmetry, by assuming that all higher order
operators are generated by the exchanges of heavy
superfields with $\sim M$ masses \cite{Frogatt}.
As we will see shortly, it is possible to find a proper set of the
heavy fermions, which after their decoupling lead to the needed
{\em rank-1} pattern of the higher order operators
fulfilling eq. (\ref{product}), and thus providing
the following Yukawa matrices at the GUT scale:
\begin{eqnsystem}{sys:ude}
&&\bordermatrix{& u^c_1 & u^c_2 & ~u^c \cr
q_1 & 0 & \eps_\Sigma\eps_H^3 c_\sigma^2 U & ~0 \cr
q_2 & \eps_\Sigma\eps_H^3 c_\sigma^2 U' & \eps_H^2 s_\sigma C & ~0 \cr
q   & 0 & 0 & ~G \cr} \cdot h_2 \\
&&
\bordermatrix{& d^c_1&d^c_2 & d^c_3 \cr
q'_1  & J\eps_\Sigma^2\eps_H^3 c_\sigma^3 D_{1}
      & J\eps_\Sigma^2\eps_H^3 c_\sigma^3 D_{2}
      & J\eps_\Sigma^2\eps_H^3 c_\sigma^3 D_{3} \cr
q'_2  & J\eps_\Sigma^2\eps_H^3 c_\sigma^3 D'_{2}
      & K\eps_\Sigma\eps_H^2 c_\sigma s_\sigma S_2
      & K\eps_\Sigma\eps_H^2 c_\sigma s_\sigma S_3 \cr
q     & 0 & 0 & \eps_H^2 c_\sigma B \cr} \cdot h_1 \\
&&
\bordermatrix{& l_1 & l_2 & l_3 \cr
e'^c_1 & \eps_\Sigma^2\eps_H^3 c_\sigma^3 D_{1}
       & \eps_\Sigma^2\eps_H^3 c_\sigma^3 D_{2}
       & \eps_\Sigma^2\eps_H^3 c_\sigma^3 D_{3}  \cr
e'^c_2 & \eps_\Sigma^2\eps_H^3 c_\sigma^3 D'_{2}
       & \eps_\Sigma\eps_H^2 c_\sigma s_\sigma S_2
       & \eps_\Sigma\eps_H^2 c_\sigma s_\sigma S_3 \cr
e^c    & 0 & 0 & \eps_H^2 c_\sigma B \cr} \cdot h_1
\end{eqnsystem}
(notice, that the basis of down quarks in $15'_{1,2}$ is already
`Cabibbo' rotated with respect to the one of the upper quarks
$15_{1,2}$ by the angle $\theta$), where $J$ and $K$ are some Clebsch
factors. As we see below, the heavy fermion mechanism leads also to
the specific predictions for the coefficients $J$ and $K$
distinguishing the down quark and charged lepton masses.

Let us introduce the set of heavy vectorlike superfields (in the
following referred as $F$-fermions) with $\sim M$ masses and
transformation properties under $SU(6)\times Z_3$ given in Table 1.
Certainly, we prescribe negative matter parity to all of them.


\begin{table}[b]
\begin{center}\begin{tabular}{c|c|c|l}
$Z_3$: & Higgs & fermions & ~~~~ $F$-fermions \\ \hline
$\omega$ & $\Sigma_1$ & $\bar{6}_i$, $\bar{6}'_i$, $20$ & $\ov{15}^2_F$,~
$\ov{15}^3_F$,~ $20_F$,~ $35_F$,~ $\ov{70}_F$,~ $84_F$ \\ \hline
$\bar{\omega}$ & $\Sigma_2$ & $15_i$ & $15^2_F$,~ $15^3_F$,~
$\ov{20}_F$,~ $\ov{35}_F$,~ $70_F$,~ $\ov{84}_F$ \\ \hline
{\em inv.} & $H$, $\bar{H}$, $Y$ & -- & $\overline{15}^1_F$,~
$15^1_F$,~ $20_F^{1,2}$,
$\overline{105}_F$, $105_F$, $\overline{210}_F$, $210_F$
\end{tabular}

\caption{$Z_3$-transformations of various supermultiplets.}
\end{center}\end{table}


Then the operators ${\cal B}$, ${\cal C}$ and ${\cal S}$ are uniquely
generated by $F$-fermion exchanges shown in Fig.~1,
with the {\em rank-1} coupling matrices (\ref{product}) in
${\cal C}$ and ${\cal S}$. Indeed, operator ${\cal B}$
defines the $\ov{6}_3$ state.
On the other hand,
the coupling with $20_F$ defines the $15_2$ state, so that the
operator ${\cal C}$ induces only the $c$ quark mass.
The coupling $(G_1 15_1 + G_2 15_2)\Sigma_1 \ov{15}^1_F$ defines the state
$15'_2=c_\theta 15_1 + s_\theta 15_2$ with $\tan\theta=G_1/G_2$,
and the couplings of $15_F^2$ define the $\ov{6}_2$ state.
Thus, the operator ${\cal S}$ induces only the $s$ and
$\mu$ masses, and in general leads to the large Cabibbo mixing.
It acts as ${\cal S} \propto
{\cal S}_1+2{\cal S}_2$, where ${\cal S}_{1,2}$ are the two possible
combinations in (\ref{dim4_smu}), so that $S^{(2)}_{ik}=2S^{(1)}_{ik}$.
Then eq. (\ref{Yuk_mat}) leads to $K=-1/5$.

The exchange of $35_F$ and $\ov{35}_F$ induces the operator ${\cal N}$
relevant for the neutrino mass (see Fig.~2). Clearly, only one combination
of neutrino states gets small Majorana mass in this way, since
$N_{kl}$ in eq. (\ref{dim4_nu}) appears to be {\em rank-1} matrix.
Then neutrino oscillations are described by one large
mixing angle.

Finally, operators ${\cal D, U}$ are generated from the $F$-fermion
exchanges shown in Fig.~3.
The operator ${\cal D}$ built in this way acts as
${\cal D} \propto {\cal D}_1+{\cal D}_3-{\cal D}_4$ with
${\cal D}_{1,2,3,4}$ being the possible convolutions in
eq. (\ref{dim5_de}). According to eq. (\ref{Yuk_de}) this leads to
$J=8/5$.
On the other hand, the operator ${\cal U}$ built as in Fig.~3,
can only mix $15_1$ state containing $u$ quark, with $15_2$ state
containing $c$ quark, but cannot provide direct mass term for the
former.\footnote{In fact, by removing the $F$-fermions $20^{1,2}_F$
one could leave the $u$ quark massless.
Though this possibility is somewhat dubious, it would naturally
solve the strong CP-problem without invoking an axion. }
As a result, the higher order operators obtained by the exchange of
$F$-fermions given in Table~1, unambiguously reproduce the
ansatz given in eqs. (\ref{sys:ude}), with $J=8/5$ and $K=-1/5$.


Before adressing the obtained fermion mass and mixing
pattern, let us remark that actually our choice of
the $F$-fermion content is a result of a rather general analysis.
In constructing the higher order operators
we have taken into account the following constraints:

(A) In order to ensure the {\em rank-1} form (\ref{product}) of the
coupling matrices,
each of the $d=6$ operators ${\cal C}$,${\cal S}$ should be
induced by the unique exchange chain.

(B) Once the exchanges generating ${\cal C}$ and ${\cal S}$ are
selected, the $d=7$ operators ${\cal D}$ and ${\cal U}$ should be
constructed by the exchange chains which are
irreducible to $d=6$ operators: otherwise the mass hierarchy
between the first and second families would be spoiled.
In other words, the exchange chains should not allow to replace
$\Sigma_1\times\Sigma_1$ by $\Sigma_2$, so that the (symmetric) tensor
product $\Sigma_1\times\Sigma_1$ should effectively act as the
$189$ or $405$ representations of $SU(6)$.
This condition requires the large representations like 105, 210, etc.
to be involved into the game.

In fact, one can classify all possible exchanges satisfying the
conditions (A) and (B). In particular, besides the exchange in Fig.~3,
operator ${\cal D}$ can be induced only by few irreducible
chains involving even larger representations. These are:
\beqn{chains}
&& 15_i\,\Sigma\, [\ov{21}_F(\ov{384}_F)+21_F(384_F)]\,\Sigma\,
[\ov{315}_F+315_F]\,\bar{H}[\ov{120}_F+120_F]\,\Sigma\,\bar{6}_k
\nonumber \\
&& 15_i\,\Sigma\, [\ov{384}_F+384_F]\,\bar{H}\,
[\ov{840}_F(\ov{1260}_F)+840_F(1260_F)]\,\Sigma\,
[\ov{84}_F(\ov{120}_F)+84_F(120_F)]\,\Sigma\,\bar{6}_k \nonumber \\
&& 15_i\,\Sigma\, [\ov{384}_F+384_F]\,\bar{H}\,[\ov{840}_F+840_F]\,
\Sigma\,[\ov{120}_F+120_F]\,\Sigma\,\bar{6}_k
\eeqn
where $\Sigma$ can be either $\Sigma_1$ or $\Sigma_2$.
%
%
These exchanges induce ${\cal D}$ respectively in the
combinations ${\cal D}_1-{\cal D}_2+{\cal D}_3+{\cal D}_4$: $J=1$,
${\cal D}_1\mp {\cal D}_4$: $J=1$,
${\cal D}_1-2{\cal D}_2-{\cal D}_4$: $J=11/17$, and thus
they all lead to unacceptable situation $\lambda_d\leq\lambda_e$.
Hence, $J=8/5$ is selected as the only one feasible choice.

One can also classify the exchanges inducing the operator ${\cal S}$.
By scanning the relevant representations for the $F$-fermions,
we have obtained that ${\cal S}$ can appear only in the combinations
${\cal S}_1$: $K=1$, ${\cal S}_2$: $K=-1/2$,
${\cal S}_1\pm {\cal S}_2$: $K=0,-2$ respectively,
${\cal S}_1-2{\cal S}_2$: $K=-1$, and
${\cal S}_1+2{\cal S}_2$: $K=-1/5$. We have chosen the latter case
uniquely selected by the exchange in Fig.~1.
All other cases are unacceptable: $K=0$ ($|K|\geq 1$) leads to massless
(or too heavy) $s$ quark,
while $K=-1/2$ \cite{BDSBH} in combination with $J=8/5$
leads to unacceptably small $m_d/m_s$ ($\approx 1/70$).

Thus, among all possible exchanges only the selected ones lead to
acceptable pattern for ${\cal D}$ and ${\cal S}$. As for the
operators ${\cal C}$ and ${\cal U}$, the only possible exchanges obeying
conditions (A) and (B) are the ones given in Figs. 1,3.

%

Let us now analyse the obtained pattern of the Yukawa matrices
(\ref{sys:ude}).
The Yukawa coupling eigenvalues and CKM weak mixing matrix
at the GUT scale are the following:
\beqn{Yuk_pattern}
 3^{rd} ~{\rm family}:~~~~ & \lambda_t=G\sim 1,
&\lambda_\tau=\lambda_b=\eps_H^2 c_\sigma B \nonumber \\
 2^{nd} ~{\rm family}:~~~~ & \lambda_c=\eps_H^2 s_\sigma C,
&\lambda_\mu=-5\lambda_s =\eps_\Sigma\eps_H^2
c_\sigma s_\sigma S_2  \nonumber \\
 1^{st} ~{\rm family}:~~~~ & \lambda_u=\eps_\Sigma^2\eps_H^4
c_\sigma^4 s_\sigma^{-1} (U'U/C),
&\lambda_e= (5/8)\lambda_d = \eps_\Sigma^2\eps_H^3 c_\sigma^3 D_{1}
\eeqn
(where the small corrections due to the mixing terms are neglected) and
\be{CKM}
V_{\rm CKM}\approx
\matr{ 1 }{ s_{12} }{ s_{12}s_{23} - s_{13}e^{-i\delta} }
{ -s_{12} }{ 1 }{ s_{23} + s_{12}s_{13}e^{-i\delta} }
{ s_{13}e^{i\delta} }{ -s_{23} }{ 1 }, ~~~~
s_{12}\approx \frac{G_1}{G_2},~~
s_{23}\approx \frac{S_3\lambda_s}{S_2\lambda_b}, ~~
s_{13}\approx \frac{D_{3}\lambda_d}{D_{1}\lambda_b}
\ee
where $\delta$ is the CP-phase.
%
In order to confront these Yukawa constants to the masses of the quarks
and leptons, we have to account for the renormalization group running.
For the heavy quarks $f=t,b,c$
we take the values of their running masses at $\mu=m_f$, while for the
light quarks $f=s,d,u$ at $\mu=1\,$GeV.
Then we have \cite{DHR,Barger}:
\beqn{masses}
&&
m_t=165\pm 15\,{\rm GeV} = A_u \eta_t y^6 \lambda_t v\sin\beta \nonumber \\
&&
m_b=4.25\pm 0.10\,{\rm GeV}= A_d \eta_b y \lambda_\tau v\cos\beta ,~~~~
m_\tau=1.784\,{\rm GeV}= A_e \eta_\tau \lambda_\tau v\cos\beta \nonumber \\
&&
m_c=1.27\pm0.05\,{\rm GeV}= A_u \eta_c y^3 \lambda_c v\sin\beta \nonumber \\
&&
m_s=100-250\,{\rm MeV}= A_d \eta_s K\lambda_\mu v\cos\beta,~~~~
m_\mu=105.6\,{\rm MeV}= A_e \eta_\mu \lambda_\mu v\cos\beta \nonumber \\
&&
m_u= (0.4\pm 0.4)m_d = A_u \eta_u y^3 \lambda_u v\sin\beta \nonumber \\
&&
m_d=(0.05\pm 0.01)m_s = A_d \eta_d J\lambda_e v\cos\beta ,~~~~
m_e=0.51\,{\rm MeV}= A_e \eta_e \lambda_e v\cos\beta
\eeqn
where $v=174\,$GeV,
\be{y}
y=\exp\left[-\frac{1}{16\pi^2}\int_{\ln m_t}^{\ln M_X}
\lambda_t^2(\mu)\mbox{d}(\ln \mu) \right]
\ee
and,
for $\alpha_s(M_Z)=0.11-0.125$
\beqn{RG_factors}
&&
\eta_b=1.5-1.6,~~~\eta_c=1.8-2.3,~~~
\eta_{s,d,u}=2.1-2.8,~~~\eta_{\tau,\mu,e}=0.99 \nonumber \\
&&
A_u=3.3-3.8, ~~~ A_d=3.2-3.7,~~~ A_e=1.5
\eeqn

It is well-known that the $b-\tau$ Yukawa unification and moderate
$\tan\beta$, both implied in our scheme, require rather large $\lambda_t$
($\lambda_t\geq 2$, so that $y=0.75-0.6$).
Then the top `pole' mass is given by its infrared
fixed limit \cite{IRfixed}
\be{Top}
M_t= m_t\left[1+\frac{4}{3\pi}\alpha_3(m_t)\right]=
(190-210)\sin\beta~ {\rm GeV}= 140-210~ {\rm GeV}
\ee
in agreement with the CDF result $M_t=174 \pm 10\pm 13\,$GeV \cite{CDF}.
Clearly, in our model $\tan\beta$ should be rather moderate:
$\tan\beta=1.2-2$. Interestingly, this range is also favoured by the
electroweak symmetry radiative breaking picture in the presence of
$b-\tau$ Yukawa unification. It is worth to mention the stricking
correlation between $M_t$ and the mass of lightest Higgs boson $M_h$.
As far as $M_t$ appears to be in the infrared fixed regime, this
correlation is essentially determined by the value of $\tan\beta$,
providing strong upper limit on $M_h$ for the low values of the latter
(see \cite{BDSBH} and refs. therein).

Then the experimental values
of $m_\tau$ and $m_c/m_\tau$ respectively imply that
$\eps_H^2 c_\sigma B \simeq 10^{-2}$
and $(C/B)\tan\sigma\tan\beta \simeq 0.4-0.6$.
{}From $m_\mu/m_\tau$ and $m_e/m_\mu$ we obtain
$\eps_\Sigma s_\sigma (S_2/B)\simeq 0.06$ and
$\eps_\Sigma\eps_H c_\sigma^2 s_\sigma^{-1} (D_1/S_2)\simeq
5\cdot 10^{-3}$. The CKM mixing pattern $|V_{us}|=0.22$,
$|V_{cb}|=0.04\pm 0.01$ and $|V_{ub}/V_{cb}|=0.1\pm 0.05$
implies respectively $G_2/G_1\simeq 4$, $S_3/S_2\simeq 3$
and $D_3/D_1\simeq 3-4$.
Taking all these into the account,
we see that our scheme gives an elegant understanding of {\em all} fermion
masses and their mixing in terms of small ratios
$\eps_H,\eps_\Sigma\sim 0.1$ and of the $O(1)$ parameters
$G,B\dots$ and $\tan\sigma$.

Moreover, we obtain
the relations
$\lambda_d=\frac{1}{5}\lambda_\mu$ and
$\lambda_d=\frac{8}{5}\lambda_e$, with possible $\sim \eps_\Sigma$
corrections that can arise
due to mixing terms in (\ref{sys:ude}).
Thus,  we have
\be{d/s}
\frac{m_d}{m_s}\simeq 8\,\frac{m_e}{m_\mu}\approx \frac{1}{25}\,
[1 + O(\eps_\Sigma)]
\ee
while for the quark running masses at $\mu=1\,$GeV we obtain
\be{s_mass}
m_s = \frac{1}{5}\,\frac{A_d\eta_s}{A_e\eta_\mu}\,m_\mu=
100-150\,\mbox{MeV},~~~~~
m_d = \frac{8}{5}\,\frac{A_d\eta_d}{A_e\eta_e}\,m_e=
4-7\,\mbox{MeV}
\ee



\vspace{0.9cm}
{\large \bf 5. Discussion }
\vspace{5mm}

As we have seen above, the fermion mass and mixing pattern
can be naturally explained in our scheme without appealing to
any horizontal symmetry, provided that the scales $M$, $V_H$ and
$V_\Sigma$ are related as $V_\Sigma/V_H \sim V_H/M \sim 0.1$.
As far as  the scale
$V_\Sigma\simeq 10^{16}\,$GeV is fixed by the $SU(5)$ unification
of gauge couplings, these relations in turn imply that
$V_H\sim 10^{17}\,$GeV and $M \sim 10^{18}\,$GeV, so that $M$ is
indeed close to the string or Planck scale.
On the other hand, the superpotential (\ref{superpot}) includes mass
parameters $M_\Sigma$ and $M_H$ which are not related to the large
scale $M$ and thus the origin of this hierarchy remains unclear.
However, bearing in mind the possibility that
our $SU(6)$ theory could be a stringy
SUSY GUT, one can assume that the superfields $H,\bar H$ and
$\Sigma_{1,2}$ are zero modes, and the Higgs superpotential has
the form not containing their mass terms:
\be{new_W}
W= \lambda Y(\bar H H -\Lambda^2) + \lambda_1\Sigma_1^3 +
\lambda_2\Sigma_2^3 + \frac{(\bar H H)}{M}\,(\Sigma_1\Sigma_2)
\ee
The last term can be effectively obtained by exchange of
the singlet superfield $Z$ with a large mass $M$,
as shown in Fig.~4. More explicitly, the relevant superpotential
has the form
\be{with_Z}
\lambda_1\Sigma_1^3 + \lambda_2\Sigma_2^3 +
\lambda Y \Sigma_1 \Sigma_2 + \lambda' Z \Sigma_1\Sigma_2 +
\rho Y(\bar H H -\Lambda^2) + \rho' Z\bar H H +
M Z^2 + M' Y^2 + \dots
\ee
(obviously, the basis of two singlets always can be redefined so that
only one of them, namely $Y$ has a linear term).
Then the relation $V_\Sigma/V_H \sim V_H/M=\eps_H$
follows naturally. Certainly, the origin of small linear
term ($\Lambda=\eps_H M$) in (\ref{new_W}) remains unclear.
It may arise due to some hidden sector outside the GUT.

Let us conclude with the following remark. In our scheme all the higher
order operators are induced by exchanges of the heavy particles
with masses $\sim M$. In doing so, all higher order operators are
under controll and the unwanted higher order operators can be always
suppressed by the proper choice of the heavy particle content.
However, the higher order operators scaled by inverse powers of the
Planck mass could appear also due to non-perturbative effects,
in an uncontrollable way.
If all such operators unavoidably occur, this would
spoil the GIFT picture. For example, already the operator
$\frac{1}{M_{Pl}} (\bar H \Sigma_1)(\Sigma_2 H)$ would provide
an unacceptably large ($\sim V_H^2/M_{Pl}$) mass to the Higgs
doublets. One may hope,
however, that not all possible structures appear in higher
order terms. Alternatively, one could try to suppress dangerous
high order operators by symmetry reasons, in order to achieve
a consistent 'all order' solution. Some possibilities
based on additional discrete (or R-type discrete) symmetries
are suggested in \cite{BCL}.

\vspace{0.9cm}
{\large \bf Acknowledgements}
\vspace{0.5cm}

This work was initiated in discussions with Riccardo Barbieri.
I would like to thank also Gia Dvali and Oleg Kancheli for discussions,
and J. Erler, L. Iba\~nez, K.S. Narain, H.P. Nilles and M. Peskin
for useful comments.



\newpage
\baselineskip=12pt

\begin{figure}\setlength{\unitlength}{1.3cm}
\begin{center}\begin{picture}(6,5.7)(0,-6)
\put(-1,-0){\makebox(0,0){ ${\cal B}$: }}
\put(-1,-2){\makebox(0,0){ ${\cal C}$: }}
\put(-1,-4){\makebox(0,0){ ${\cal S}$: }}
\put(0,0){\line(1,0){6}}
\multiput(1,0)(2,0){3}{\line(0,-1){1}}
\put(0,0.1){\makebox(0,0)[bl]{20}}
\put(2,0.1){\makebox(0,0)[b]{$\overline{70}_{\rm F}~~~70_{\rm F}$}}
\put(4,0.1){\makebox(0,0)[b]{$\overline{15}^3_{\rm F}~~~15^3_{\rm F}$}}
\put(6,0.1){\makebox(0,0)[br]{$\bar{6}_3$}}
\put(1.1,-1){\makebox(0,0)[bl]{$\Sigma_1$}}
\put(3.1,-1){\makebox(0,0)[bl]{$\bar{H}$}}
\put(5.1,-1){\makebox(0,0)[bl]{$\bar{H}$}}
\put(2,0){\makebox(0,0){$\times$}}
\put(4,0){\makebox(0,0){$\times$}}
%
%
\put(0,-2){\line(1,0){6}}
\multiput(1,-2)(2,0){3}{\line(0,-1){1}}
\put(0,-1.9){\makebox(0,0)[bl]{$15_2$}}
\put(2,-1.9){\makebox(0,0)[b]{$20_{\rm F}~~~\overline{20}_{\rm F}$}}
\put(4,-1.9){\makebox(0,0)[b]{$\overline{20}_{\rm F}~~~20_{\rm F}$}}
\put(6,-1.9){\makebox(0,0)[br]{$15_2$} }
\put(1.1,-3){\makebox(0,0)[bl]{$H$} }
\put(3.1,-3){\makebox(0,0)[bl]{$\Sigma_2$}}
\put(5.1,-3){\makebox(0,0)[bl]{$H$}}
\put(2,-2){\makebox(0,0){$\times$}}
\put(4,-2){\makebox(0,0){$\times$}}
\put(0,-4){\line(1,0){6}}
\multiput(1,-4)(2,0){3}{\line(0,-1){1}}
\put(0,-3.9){\makebox(0,0)[bl]{$15'_2$} }
\put(2,-3.9){\makebox(0,0)[b]{$\overline{15}^1_{\rm F}~~~15^1_{\rm F}$}}
\put(4,-3.9){\makebox(0,0)[b]{$\overline{15}^2_{\rm F}~~~15^2_{\rm F}$}}
\put(6,-3.9){\makebox(0,0)[br]{$\bar{6}_{2,3}$}}
\put(1.1,-5){\makebox(0,0)[bl]{$\Sigma_1$}}
\put(3.1,-5){\makebox(0,0)[bl]{$\Sigma_2$}}
\put(5.1,-5){\makebox(0,0)[bl]{$\bar{H}$}}
\put(2,-4){\makebox(0,0){$\times$}}
\put(4,-4){\makebox(0,0){$\times$}}
\end{picture}
\caption{diagrams giving rise\label{Diagrams}
to the operators ${\cal B,~ C,~ S}$ respectively.}
\end{center}\end{figure}
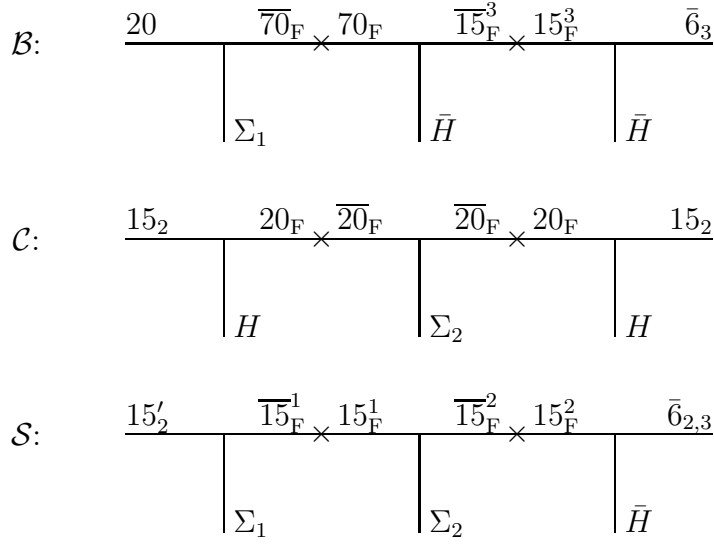


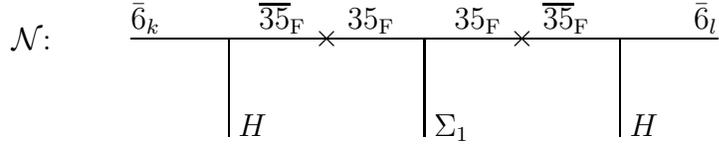
\begin{figure}\setlength{\unitlength}{1.3cm}
\begin{center}\begin{picture}(6,2)(0,-1)
\put(-1,-0){\makebox(0,0){ ${\cal N}$: }}
\put(0,0){\line(1,0){6}}
\multiput(1,0)(2,0){3}{\line(0,-1){1}}
\put(0,0.1){\makebox(0,0)[bl]{$\bar{6}_k$}}
\put(2,0.1){\makebox(0,0)[b]{$\ov{35}_{\rm F}~~~~35_{\rm F}$}}
\put(4,0.1){\makebox(0,0)[b]{$35_{\rm F}~~~~\ov{35}_{\rm F}$}}
\put(6,0.1){\makebox(0,0)[br]{$\bar{6}_l$}}
\put(1.1,-1){\makebox(0,0)[bl]{$H$}}
\put(3.1,-1){\makebox(0,0)[bl]{$\Sigma_1$}}
\put(5.1,-1){\makebox(0,0)[bl]{$H$}}
\put(2,0){\makebox(0,0){$\times$}}
\put(4,0){\makebox(0,0){$\times$}}
\end{picture}\caption{the diagram
giving rise to the operator ${\cal N}$ for neutrino mass .\label{FalseMass}}
\end{center}\end{figure}


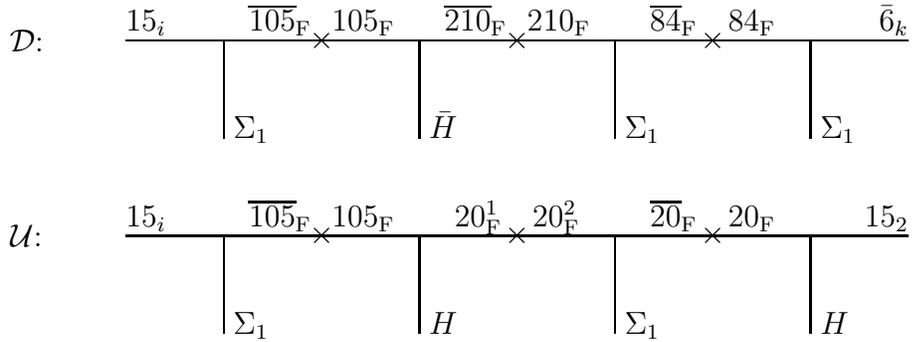
\begin{figure}\setlength{\unitlength}{1.3cm}
\begin{center}\begin{picture}(6,4)(0,-3.3)
\put(-1,-0){\makebox(0,0){ ${\cal D}$: }}
\put(-1,-2){\makebox(0,0){ ${\cal U}$: }}
%
\put(0,0){\line(1,0){8}}
\multiput(1,0)(2,0){4}{\line(0,-1){1}}
\put(0,0.1){\makebox(0,0)[bl]{$15_{i}$}}
\put(2,0.1){\makebox(0,0)[b]{$\overline{105}_{\rm F}~~105_{\rm F}$}}
\put(4,0.1){\makebox(0,0)[b]{$\overline{210}_{\rm F}~~210_{\rm F}$}}
\put(6,0.1){\makebox(0,0)[b]{$\overline{84}_{\rm F}~~~84_{\rm F}$}}
\put(8,0.1){\makebox(0,0)[br]{$\bar{6}_{k}$}}
\put(1.1,-1){\makebox(0,0)[bl]{$\Sigma_1$} }
\put(3.1,-1){\makebox(0,0)[bl]{$\bar{H}$} }
\put(5.1,-1){\makebox(0,0)[bl]{$\Sigma_1$} }
\put(7.1,-1){\makebox(0,0)[bl]{$\Sigma_1$} }
\put(2,0){\makebox(0,0){$\times$}}
\put(4,0){\makebox(0,0){$\times$}}
\put(6,0){\makebox(0,0){$\times$}}
\put(0,-2){\line(1,0){8}}
\multiput(1,-2)(2,0){4}{\line(0,-1){1}}
\put(0,-1.9){\makebox(0,0)[bl]{$15_{i}$}}
\put(2,-1.9){\makebox(0,0)[b]{$\overline{105}_{\rm F}~~105_{\rm F}$}}
\put(4,-1.9){\makebox(0,0)[b]{$20^1_{\rm F}~~~20^2_{\rm F}$}}
\put(6,-1.9){\makebox(0,0)[b]{$\overline{20}_{\rm F}~~~20_{\rm F}$}}
\put(8,-1.9){\makebox(0,0)[br]{$15_2$}}
\put(1.1,-3){\makebox(0,0)[bl]{$\Sigma_1$} }
\put(3.1,-3){\makebox(0,0)[bl]{$H$} }
\put(5.1,-3){\makebox(0,0)[bl]{$\Sigma_1$} }
\put(7.1,-3){\makebox(0,0)[bl]{$H$} }
\put(2,-2){\makebox(0,0){$\times$}}
\put(4,-2){\makebox(0,0){$\times$}}
\put(6,-2){\makebox(0,0){$\times$}}
%
%
\end{picture}
\caption{diagrams giving rise\label{Diagrams2}
to the operators ${\cal D}$ and ${\cal E}$ respectively.}
\end{center}\end{figure}



\begin{figure}\setlength{\unitlength}{1.5mm}
\begin{center}\begin{picture}(40,20)
\put(0,0){\vector(1,1){5}}   \put(5,5){\line(1,1){5}}
\put(0,20){\vector(1,-1){5}}\put(5,15){\line(1,-1){5}}
\put(10,10){\line(1,0){20}}
\put(40,20){\vector(-1,-1){5}} \put(35,15){\line(-1,-1){5}}
\put(40,0){\vector(-1,1){5}}\put(35,5){\line(-1,1){5}}
\put(15,10){\vector(-1,0){0}}\put(15,11){\makebox(0,0)[b]{$Z$}}
\put(25,10){\vector(1,0){0}}\put(25,11){\makebox(0,0)[b]{$Z$}}
\put(6,15){\makebox(0,0)[bl]{$\Sigma_1$}}
\put(6,5){\makebox(0,0)[tl]{$\Sigma_2$}}
\put(34,15){\makebox(0,0)[br]{$H$}}
\put(34,5){\makebox(0,0)[tr]{$\bar H$}}
\put(20,10){\makebox(0,0){$\times$}}
\end{picture}
\caption{Diagram generating the operator
$\frac{1}{M}(\bar H H)(\Sigma_1\Sigma_2)$.\label{pDecDiagram}}
\end{center}\end{figure}
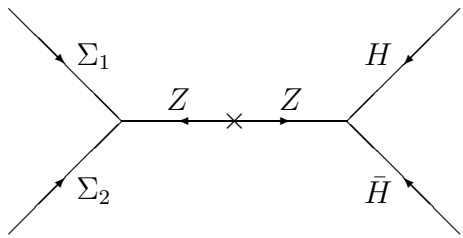


\end{document}